\documentstyle[12pt]{article}
\begin{document}
\def\subsubsection{\@startsection{subsubsection}{3}{\z@}{-3.25ex plus
 -1ex minus -.2ex}{1.5ex plus .2ex}{\large\sc}}
\renewcommand{\thesection}{\arabic{section}}
\renewcommand{\thesubsection}{\arabic{section}.\arabic{subsection}}
\renewcommand{\thesubsubsection}
{\arabic{section}.\arabic{subsection}.\arabic{subsubsection}}
\pagestyle{plain}
\sloppy
\textwidth 150mm
\textheight 620pt
\topmargin 35pt
\headheight 0pt
\headsep 0pt
\topskip 1pt
\oddsidemargin 0mm
\evensidemargin 10mm
\setlength{\jot}{4mm}
\setlength{\abovedisplayskip}{7mm}
\setlength{\belowdisplayskip}{7mm}
\newcommand{\be}{\begin{equation}}
\newcommand{\bel}[1]{\begin{equation}\label{#1}}
\newcommand{\ee}{\end{equation}}
\newcommand{\bea}{\begin{eqnarray}}
\newcommand{\ba}{\begin{array}}
\newcommand{\eea}{\end{eqnarray}}
\newcommand{\ea}{\end{array}}
\newcommand{\noin}{\noindent}
\newcommand{\ra}{\rightarrow}
\newcommand{\txs}{\textstyle}
\newcommand{\disp}{\displaystyle}
\newcommand{\scs}{\scriptstyle}
\newcommand{\scscs}{\scriptscriptstyle}
\newcommand{\sx}[1]{\sigma^{\, x}_{#1}}
\newcommand{\sy}[1]{\sigma^{\, y}_{#1}}
\newcommand{\sz}[1]{\sigma^{\, z}_{#1}}
\newcommand{\sP}[1]{\sigma^{\, +}_{#1}}
\newcommand{\sM}[1]{\sigma^{\, -}_{#1}}
\newcommand{\spm}[1]{\sigma^{\,\pm}_{#1}}
\newcommand{\Spm}{S^{\,\pm}}
\newcommand{\Sz}{S^{\, z}}
\newcommand{\hspf}{\hspace*{5mm}}
\newcommand{\hspt}{\hspace*{2mm}}
\newcommand{\vspf}{\vspace*{5mm}}
\newcommand{\vspt}{\vspace*{2mm}}
\newcommand{\hsix}{\hspace*{6mm}}
\newcommand{\hfour}{\hspace*{4mm}}
\newcommand{\vsix}{\vspace*{6mm}}
\newcommand{\vfour}{\vspace*{4mm}}
\newcommand{\vtwo}{\vspace*{2mm}}
\newcommand{\htwo}{\hspace*{2mm}}
\newcommand{\honecm}{\hspace*{1cm}}
\newcommand{\vonecm}{\vspace*{1cm}}
\newcommand{\htwocm}{\hspace*{2cm}}
\newcommand{\vtwocm}{\vspace*{2cm}}
\newcommand{\ru}{\rule[-2mm]{0mm}{8mm}}
\newcommand{\tinf}{\rightarrow \infty}
\newcommand{\udl}{\underline}
\newcommand{\ovl}{\overline}
\newcommand{\nwl}{\newline}
\newcommand{\nwp}{\newpage}
\newcommand{\clp}{\clearpage}
\newcommand{\simleq}{\raisebox{-1.0mm}
{\mbox{$\stackrel{\textstyle <}{\sim}$}}}
\newcommand{\simgeq}{\raisebox{-1.0mm}
{\mbox{$\stackrel{\textstyle >}{\sim}$}}}
\newcommand{\half}{\mbox{\small$\frac{1}{2}$}}
\newcommand{\smfrac}[2]{\mbox{\small$\frac{#1}{#2}$}}
\newcommand{\bra}[1]{\mbox{$\langle \, {#1}\, |$}}
\newcommand{\ket}[1]{\mbox{$| \, {#1}\, \rangle$}}
\newcommand{\exval}[1]{\mbox{$\langle \, {#1}\, \rangle$}}
\newcommand{\BIN}[2]
{\renewcommand{\arraystretch}{0.8}
\mbox{$\left(\ba{@{}c@{}}{\scs #1}\\{\scs #2}\ea\right)$}
\renewcommand{\arraystretch}{1}}
\newcommand{\UQSU}{\mbox{U$_{q}$[sl(2)]}}
\newcommand{\UQSUA}{\mbox{U$_{q}$[$\widehat{\mbox{sl(2)}}$]}}
\newcommand{\comm}[2]{[\, #1 , \, #2 \, ]}
\newcommand{\intpa}[1]{\mbox{\small $[ \, #1 \, ]$}}
\newcommand{\BC}{boundary conditions}
\newcommand{\TBC}{toroidal boundary conditions}
\newcommand{\ABC}{antiperiodic boundary conditions}
\newcommand{\PBC}{periodic boundary conditions}
\newcommand{\CBC}{cyclic boundary conditions}
\newcommand{\OBC}{open boundary conditions}
\newcommand{\XXZ}{XXZ Heisenberg chain}
\newcommand{\ZFQC}{Zamolodchikov Fateev quantum chain}
\newcommand{\FSS}{finite-size scaling}
\newcommand{\FSSS}{finite-size scaling spectra}
\newcommand{\DT}{duality transformation}
\newcommand{\POM}{Potts models}
\newcommand{\PM}{projection mechanism}
\newcommand{\PQC}{Potts quantum chain}
\newcommand{\TLA}{Temperley-Lieb algebra}
\newcommand{\PTLA}{periodic Temperley-Lieb algebra}
\newcommand{\HA}{Hecke algebra}
\newcommand{\VA}{Virasoro algebra}
\newcommand{\CFT}{conformal field theory}
\newcommand{\SS}{steady state}
\newcommand{\FS}{Fock space}
\newcommand{\FSF}{Fock space formalism}
\newcommand{\ME}{master equation}
\newcommand{\DEX}{diffusion with particle exclusion}
\newcommand{\EVOP}{evolution operator}
\newcommand{\Gtwo}{two-point correlation function}

\begin{titlepage}
\thispagestyle{empty}
\begin{center}
\vspace*{1cm}
{\Large \bf
Non-equilibrium Dynamics of Finite Interfaces
}\\[30mm]

{\Large D. B. Abraham, T. J. Newman and G. M. Sch{\"u}tz} \\[2cm]

\begin{minipage}[t]{13cm}
\begin{center}
{\small\sl
Department of Physics, University of Oxford, \\
Theoretical Physics, 1 Keble Road, OX1 3NP, U.K.
}\end{center}
\end{minipage}
\vspace{28mm}
\end{center}
{\small
We present an exact solution to an interface model representing the dynamics
of a domain wall in a two-phase Ising system. The model is microscopically
motivated, yet we find that in the scaling regime our results are consistent
with those obtained previously from a phenomenological, coarse-grained Langevin
approach.}
\\
\vspace{5mm}\\
\udl{PACS numbers:} 75.40.Gb, 05.50.+q, 05.70.Ln
\end{titlepage}

\newpage
\baselineskip 0.3in

The study of the magnetisation profile between coexistent phases in models
of subcritical uniaxial ferromagnets (and their analogues) has generated
important and unexpected theoretical advances. A typical example is the
existence of spatial fluctuations of the interface location so large that
they diverge with system size. Phenomenologically speaking, these fluctuations
are generated by capillary waves described in a continuum theory of an
interface which is a sharp dividing surface between oppositely magnetised
phases, controlled by a somewhat arbitrary short range spatial cutoff
\cite{1}. Against this, one has to set the density functional theories
going back to van der Waals and Maxwell \cite{2} which do not give divergent
spatial
interface fluctuations. It is fortunate that both exact and rigorous results
are available for the planar Ising model which resolve the conflict in favour
of the large spatial fluctuations \cite{3}, but have yet to resolve the
problem of local interface structure. The conclusions of the exact calculation
were recaptured in a Helmholtz fluctuation theory for the phase separating
surface \cite{4} which brings in the concept of interfacial stiffness and
which is valid in a spatially coarse-grained sense, much as in Widom scaling
theory \cite{5}. These fluctuations are an essential ingredient in the
statistical mechanics of a number of surface phase transition phenomena
\cite{6}. In this letter we focus on dynamical aspects of these fluctuations
by studying an exactly solvable model on a finite lattice.

First, we recall an
equilibrium result which helps motivate the model itself
and an earlier phenomenological treatment. Suppose in a zero-field planar
Ising ferromagnetic model, the interface between coexistent phases is
established and localised in laboratory-fixed axes by specifying the boundary
spins as shown in Fig. 1. The straight line of length 2L connecting the
spin-flip points is the Wulff shape for the interface. It is convenient to
define coordinates parallel and perpendicular to this line, with origin at
its centre. Let the thermodynanic limit of infinite strip length be taken
first. Then denoting the magnetisation at $(x,y)$ by $m(x,y/L)$ we have for
$-1 < \alpha < 1$ \cite{3,7}
\bel{1}
\tilde{m}(\alpha,\beta) = \lim_{L\ra\infty}
m(\alpha L, \beta L^{\delta}/L) = m^{\ast}
\Phi\left(\frac{b(\vartheta,T) \beta}{\sqrt{1-\alpha^2}}\right) \hfour
(\delta=1/2)
\ee
where $m^{\ast}$ is the spontaneous magnetisation, $\Phi(x)$ is the error
function
${\rm erf}(x)$ and
$b(\vartheta,T) = \sigma(\vartheta) +
\sigma''(\vartheta)$ is the surface stiffness. (Here $\sigma(\vartheta)$ is the
angle-dependent surface tension.) On the other hand, one has
$\tilde{m}(\alpha,\beta) = 0 \
(m^{\ast}\mbox{sign}(\beta))$ for $\delta < 1/2 \ (> 1/2)$.

In this letter we study the dynamical case with a new exact
result. In order to generalise to the dynamical case we need to define a model
with
dynamics appropriate to the physical system described above. We shall see that
our
model reproduces the equilibrium result (\ref {1}) in the infinite time scaling
limit.
To motivate our model, note that as $T \ra 0$,
$b(\vartheta) > 0$ (strictly) provided $\vartheta \neq 0, \pm \pi/2$;
for $\vartheta=0, \pm \pi/2$, $b(\vartheta,T) \ra 0$
as $T \ra 0$. This is a primitive example of facetting.
At $T=0$, the Peierls contours reduce to paths connecting $(-L,0)$  to
$(L,2L \tan{\vartheta})$ which either step to the right or upwards
($\tan{\vartheta}>0$); all such paths are degenerate.
They separate regions of opposite magnetisation; since $T=0$, this
magnetisation is of
unit magnitude. In this case, the
theorem is related to normal fluctuations in coin tossing.
Let us now concentrate on the case $\tan{\vartheta}=1$, with an initial
sawtooth configuration as shown in Fig. 2. For later convenience we
have rotated the system by 45 degrees.

At $T=0$, any minimum energy path can be represented as a sequence
$S=\{n_{-L+1},
n_{-L+2}, \dots , n_{L}\}$ of $2L$ binary numbers $n_k$ where $n_k=1$ if the
$k^{th}$ segment of the interface steps upwards (in an angle of 45 degrees)
and $n_k=0$ if the steps goes downwards (see Fig. 2). One can think of
$n_k$ as an occupation number which is related
to the interface height $h_k$ in the rotated system by  $2n_k - 1 =
h_k - h_{k-1}$. The configuration at any time $t$ is then given as
a time-slice $S(t) = \{n_k(t) : k=-L+1, \dots, L\}$ and the dynamics are
given by specifying rules connecting $S(t+1)$ to $S(t)$.

We take particle-hole exchange on neighbouring lattice sites with probability
$p$, but leave
pairs of particles or holes unchanged. This corresponds to single spin-flips
at local extrema of the interface. The nearest neighbour interaction makes
it advisable to update on alternating sublattices at subsequent time-steps.
This updating scheme turns our model into a cellular automaton which is
related to the six vertex model on a lattice oriented at 45 degrees
\cite{8,9}: take $n_k = 1$ (resp. =0) to be an upwards (resp. downwards)
pointing arrow as in Fig. 3 where we show the ice-vertices, the
corresponding dynamical events and their weights in the particle-hole
picture. With $0\leq p \leq 1$, the two row-to-row transfer matrices in
the (1,1) direction generate a stochastic time evolution satisfying local
detailed balance.

The transfer matrix can be written as
\bel{3}
T = T^{even} T^{odd} = \prod_{j=-L+1}^{L-1} T_{2j} \prod_{j=-L+1}^{L} T_{2j-1}
\ee
with $T_j=1-pe_j$, $e_j = ( \sigma_j^x\sigma_{j+1}^x +
\sigma_j^y\sigma_{j+1}^y + \sigma_j^z\sigma_{j+1}^z -1)/2$, the
$\sigma^{x,y,z}$ being the Pauli matrices. We choose $n_k = (1-\sigma_k^z)/2$,
a vacuum $|0\rangle$ with $n_k|0\rangle=0$. The initial state is the
$L$-particle state
$|v\rangle = (|1,0,1,0,\dots\rangle + |0,1,0,1,\dots\rangle)/2$. The
displacement of the interface at $(x,t)$ is thus
\bel{4}
h(x,t) = \sum_{k=-L+1}^{x} \left(1-2n_k(t)\right)
\ee
with zero mean by spin-reversal symmetry, but with
\bel{5}
w^2(x,t) = \langle\left(h(x,t)\right)^2\rangle =
4 \sum_{k,l=-L}^{x} \langle L | n_k n_l T^t |v\rangle - (x+L)^2 \htwo .
\ee
Here $\langle L|$ is the out state which sums over all eigenstates of the $n_k$
with particle number $N=L$, giving each such state equal weight. Evidently
\bel{6}
\langle L | = (L!)^{-1} \langle 0| (S^{+})^L
\ee
where $S^{\pm} = \sum s_k^{\pm}$ and $s_k^{\pm}=(\sigma_k^x\pm i\sigma_k^y)/2$.
Using the commutator
\bel{7}
\left[ (S^{+})^L \;, \; n_k \right] = L s_k^{+} (S^{+})^{L-1}
\ee
and the $SU(2)$ symmetry of $T$, (\ref{6}) and (\ref{7}) together with
(\ref{5}) give
\bel{8}
w^2(x,t) = 4 \sum_{k,l=-L+1}^{x} \sum_{p,q=-L+1}^{L} M_1(k,l;p,q|t) M_2(p,q)
+ 2(x+L) - (x+L)^2
\ee
where
\bel{9}
M_1(k,l;p,q|t) = \langle 0|s_k^{+}s_l^{+} T^t s_p^{-} s_q^{-}|0\rangle
\ee
is a matrix element in the 2-particle sector and
\bel{10}
M_2(p,q) = \langle 0| s_p^{+} s_q^{+} (S^{+})^{L-2} |v\rangle /(L-2)!
\ee
contains the information coming from the initial condition. Using (\ref{6})
and the expression for $|v\rangle$ above, it follows that $M_2(p,q) = 1/2$
for $(p-q)$ even and =0 otherwise. Thus
\bel{11}
w^2(x,t) = 2 \sum_{k,l=-L+1}^{x}
\sum_{\stackrel{p,q=-L+1}{p-q \mbox{\small even}}}^{L} M_1(k,l;p,q|t) +2(x+L)
- (x+L)^2
\ee
which is an enormous simplification of (\ref{5}).
Expressions for higher moments can be obtained in a similar way.

We study first the infinite-time limit, thus deriving the static magnetisation
$m(x,y) = 1-2\langle \Theta(h(x) - y) \rangle$ where $\Theta(s)$ is the
Heavyside step function. The (normalised) $L$-particle steady state of the
system is $| L \rangle / B(2L,L)$ with the binomial coefficient $B(m,n) =
m!/(n!(m-n)!)$. A short calculation yields
\bel{11a}
m(x,y) =  \frac{\mbox{sign}(y)}{B(2L,L)} \sum_{k=-y+1}^{y-1}
B(L-x,\frac{L-x}{2}-k)
B(L+x,\frac{L+x}{2}+k) \htwo .
\ee
In the scaling limit $x=\alpha L$ and $y=\beta L^\delta/L$ one obtains
again the scaling form (\ref{1}) with surface stiffness $b=1$ and $m^{\ast}=1$.

Now we turn to the dynamics.
For an exact calculation of the transition amplitudes (\ref{9}) one may either
derive difference equations in $t$ and $x$ as in \cite{9} or
use the Bethe ansatz \cite{10}. It simplifies matters if (a)
we pass
to a continous-time formulation by taking the limit $p \ra 0$ and
$t \ra \infty$ such that $\tau = 2tp$ remains fixed and (b) by taking
periodic boundary conditions (i.e. by studying the transfer matrix $T_P =
T_L T$ instead of $T$). Neither of these simplifications is expected to be
of physical relevance to the system \cite{8,9}.
Note, however, that $w^2(x,\tau)$ now
measures fluctuations in the height differences $h(x,\tau) - h(-L,\tau)$.
By taking this infinite time limit the evolution operator $T_P^t$ becomes
the evolution operator $\exp{(-H\tau)}$ of the symmetric simple exclusion
process \cite{11}, but defined on a {\em finite} lattice with $2L$ sites.
$H=-{1\over2}\sum_{k=-L+1}^{L} e_k$ is the quantum Hamiltonian of the isotropic
ferromagnetic Heisenberg model in one dimension.

In order to derive an exact expression for the height fluctuations (\ref{11})
as a function of $x$ and $\tau$ we define the translationally invariant states
\bel{12}
|0,r\rangle = \frac{1}{\sqrt{2L}}
\sum_{k=-L+1}^{L} s_k^{+} s_{k+r}^{+} |0\rangle
\hfour (1 \leq r \leq L) \htwo .
\ee
Note that $|r\rangle$, $|-r\rangle$ and $|r+2L\rangle$ are identical and
$\langle r|r \rangle = 1 + \delta_{r,L}$. Using translational invariance of
$H$ one can write with these conventions
\bel{13}
w^2(x,t) = 2 \sum_{r=-L+1}^{x-1} \sum_{R=1}^{L-1} (L+x-r) \langle 0,r|
\exp{(-H\tau)} |0,2R\rangle + 2 (x+L) - (x+L)^2
\ee
The matrix elements $c_R(r,\tau)=\langle 0,r|\exp{(-H\tau)}|0,R\rangle$
satisfy the differential-difference equation
\bel{14}
\ba{lcl}
\frac{\partial}{\partial \tau} c_R(r,\tau) & = &
c_R(r+1,\tau) + c_R(r-1,\tau) - 2c_R(r,\tau) \hfour (r > 1) \vtwo \\
\frac{\partial}{\partial \tau} c_R(r,\tau) & = &
c_R(r+1,\tau) - c_R(r,\tau) \hfour (r = 1)
\ea
\ee
with initial condition $c_R(r,0)=\delta_{r,R}(1+\delta_{R,L})$. We note in
passing that one may obtain these equations of motion directly from a master
equation formulation of the problem \cite{13} , indicating that the usual
stochastic formulation of the exclusion process is isomorphic to the
continuous-time limit of the model defined above.
Taking into account the
periodicity and reflection  properties of the states $|0,r\rangle$
one finds
\bel{15}
c_R(r,\tau) = e^{-2\tau} \sum_{m=-\infty}^{\infty} \left(
I_{r-R+(2L-1)m}(2\tau)
+I_{r+R-1 + (2L-1)m}(2\tau) \right)
\ee
where $I_p(x)$ are  modified Bessel functions.
With (\ref{13}) and
\bel{16}
\sum_{m=-\infty}^{\infty} I_{p+mN}(z) = \frac{1}{N}\sum_{k=0}^{N-1}
e^{z\cos{(2\pi k/N)}} \; e^{-2\pi i kp/N}
\ee
we finally obtain
\bel{17}
w^2(x,\tau) = \frac{4}{2L-1}\sum_{k=1}^{2L-1}
\left( 1-(-1)^k \frac{\cos{2\pi kx/(2L-1)}}{\cos{\pi k/(2L-1)}} \right)
\frac{1-e^{-2(1-\cos{2\pi k/(2L-1)})\tau}}{1-\cos{2\pi k/(2L-1)}}
 + \epsilon (x)
\ee
where $\epsilon (x) = 0 \ (1)$ for $x$ even (odd).

It is instructive to compare this exact result with that obtained from
a phenomenological approach based on a Langevin description \cite{14} - the
dynamics
of the interface are assumed to be described by the following additive noise
Langevin
equation
\bel{19}
\partial_{t} h_{j} = {1\over2}(h_{j-1}+h_{j+1}-2h_{j}) + \eta _{j}(t)
\ee
where $\eta _{j}(t)$ is the usual gaussian white noise.
One easily obtains
\bel{17a}
w^2(x,\tau) = \frac{4}{2L}\sum_{k=1}^{2L}
\left( 1-(-1)^k\cos{2\pi kx/(2L)} \right)
\frac{ 1-e^{-2(1-\cos{2\pi k/(2L)})\tau} }{1-\cos{2\pi k/(2L)}}
\htwo .
\ee
The similarity between the two expressions (\ref{17}) and (\ref{17a})  is
striking,
considering the vast simplifications inherent in the phenomenological Langevin
model.

It is interesting to take the scaling limit of the above expressions, i.e. we
take $x,\tau ,L \rightarrow \infty $ with $\alpha \equiv x/L \mbox{ and } u
\equiv
\tau/L^{2} $ fixed. We then find from both (\ref{17}) and (\ref{17a}).
\bel{18}
\lim \limits _{L \rightarrow \infty} {w^{2}(\alpha, u)\over 2L} =
2 \int_0^u ds \left( \theta_3(0) -
\theta _{3}(\pi (\alpha - 1)/2) \right)
\ee
where $\theta _{3}(v) \equiv \theta_3(v,q)$ is the theta function with nome
$q = \exp (-\pi^2 u)$.
We obtain the following asymptotic forms of the scaling function: for arbitrary
$\alpha $ and $u \rightarrow \infty $ we have
\bel{18a}
\lim \limits _{L \rightarrow \infty} {w^{2}(\beta, u)\over L} =
{1\over 4}(1-\alpha^{2}) - {2\over \pi^{2}}(1+\cos{\pi \alpha})e^{-\pi^{2}u}
\ee
and for $\alpha =0$ and $u \ll 1$ we have
\bel{18b}
\lim \limits _{L \rightarrow \infty} {w^{2}(0, u)\over L} = 2 \sqrt{u/\pi}
\htwo .
\ee

The fact that the scaling function evaluated from the {\em exact} dynamics is
identical
to that obtained from the Langevin description (\ref {19})
is very surprising, since the latter approach completely neglects the strong
dynamical
constraints  of the original model, i.e. the restricted possible values of
neighbouring
height differences.

We may gain some insight into this result by considering the corresponding
particle
dynamics for the Langevin description (\ref {19}). The appropriate particle
picture
is one of symmetric diffusion with {\em no} exclusion, the particles to be
interpreted
as units of height gradient. One may derive an exact Langevin equation for
this non-exclusive particle process which then may be mapped to (\ref {19}) in
the height
variables. This is done by Taylor expanding the master equation
for the distribution $P(\{ n_{j}\},t)$ (where $n_{j}$ is the {\em unrestricted}
occupation number at site $j$),
thus deriving a Fokker-Planck equation in the Stratonovich representation. The
corresponding
Langevin equation for the occupation numbers may then be mapped to the original
height
variables reproducing (\ref {19}). It is of interest to compare this to an {\em
exact}
Langevin equation for the exclusion case that we have studied in this paper.
Using the
stochastic Grassmann variables introduced in \cite {13} one may derive an exact
Langevin
equation for the `height' variables  $\lbrace f_{j} \rbrace$
of the form
\bel{22}
\partial_{t} f_{j} = {1\over2}(f_{j-1}+f_{j+1}-2f_{j})(1 + 2^{1/2}\eta _{j}(t))
\ee
(these variables are of Grassmann type, but are simply related to the original
height
variables $\lbrace h_{j} \rbrace$).
The essential difference between this exact Langevin equation and that given in
(\ref {19})
is the appearance of multiplicative noise, which appears to be irrelevant in
the scaling
regime.
Therefore the equivalence between the scaling functions for the restricted and
unrestricted
interface models may be thought of as an equivalence between symmetric
diffusion with or
without exclusion -
even for non-zero values of the scaling variable $u$.

We have yet to derive a full dynamical version of the magnetisation (\ref{1})
for our model,
as was done in \cite{14} for the Langevin dynamics with additive noise.
To this end, one has to calculate all higher even moments of the
height fluctuations. (The odd moments all vanish due to spin reversal
symmetry.)
As regards our model and the phenomenological model of Ref. \cite{14}, it is
unclear
whether the identity in the scaling forms of the height fluctuations derived
above
persists beyond the second moments.
However, having shown that in the scaling limit both the static magnetisation
$m(x,y)$
and $w^2(x,t)$ coincide, we conclude that the simplified Langevin dynamics with
additive noise represent a qualitatively and quantitatively adequate approach
to this problem in large but finite systems in the scaling region.

We note that our discussion of the dynamics is limited to the continuous time
limit
defined above. One expects no difference in the scaling form (\ref {18}) for
non-zero
hopping rate $p$ (except that the natural scaling variable is now $t/L^{2}$ as
opposed
to $\tau /L^{2}$. However, the continuous time limit defined by $(1-p)
\rightarrow 0, t
\rightarrow \infty $ with $\tau '=  t(1-p)$ fixed has qualitatively distinct
scaling
behaviour since it corresponds to a `relativistic' limit of the dynamics \cite
{9}.
The behaviour of the interface fluctuations for this case will be discussed
elsewhere
\cite {15}.

In summary, we have investigated a particular stochastic model of a finite
interface for which the dynamics act on the microscopic degrees of freedom,
rather than on the phenomenologically specified, coarse-grained microscopic
variables as is the case in Langevin theories. Nevertheless, results obtained
so
far for our model
agree with the earlier Langevin theory, encouraging its continued
use in appropriate contexts where the truly microscopic theories are not
analytically amenable.

\vskip .5in
The authors acknowledge financial support from the SERC, and the hospitality
and stimulating environment of both the Isaac Newton Institute, Cambridge and
the Weizmann Institute. They thank E. Domany, P. Ferrari, C. Kipnis and
E. Presutti for encouraging remarks.

\newpage
\bibliographystyle{unsrt}

\newpage
\noin
{\bf List of Figure Captions}

\vskip 1.in
\noin
Fig. 1: The Wulff profile seperating regions of opposite magnetisation. The
lower (upper)
region has negative (positive) magnetisation.

\vskip .5in
\noin
Fig. 2: The mapping between the restricted interface and the particle exclusion
process.
In a) we indicate the initial condition of a flat interface corresponding to
alternating
sites in the particle model being occupied. In b) we show a possible interface
configuration
at some later time and the corresponding particle occupancies. The indicated
flips in the
interface correspond to particles hopping on the lattice.

\vskip .5in
\noin
Fig. 3: Allowed vertex configurations in the six-vertex model and their
Boltzmann weights.
Up-pointing arrows correspond to particles, down-pointing arrows represent
vacant sites.
In the dynamical interpretation of the model the Boltzmann weights give the
transition
probability of the state represented by the pair of arrows below the vertex to
that above
the vertex.

\end{document}